%% file: paphep.tex
\documentclass{elsart}
\usepackage{english}
\usepackage{feynmf}
\usepackage{feynmakrot}
\usepackage{epsfig}
\def\dd#1{D_\mathrm{#1}}

\newcommand\Tr{{\rm{Tr}}}
\newcommand\bea{\eqnarray}
\newcommand\eea{\endeqnarray}
\newcommand\beq{\begin{equation}}
\newcommand\eeq{\end{equation}}
\newcommand\nn{\nonumber\\}
\begin{document}
\begin{fmffile}p
\begin{frontmatter}
\begin{flushright}
hep-ph/9702255
\end{flushright}
\title{
SU(5)+Adjoint Higgs Model at Finite Temperature}
\author{Arttu Rajantie\thanksref{osoite}}
\thanks[osoite]{E-mail: Arttu.Rajantie@Helsinki.Fi}
\address{ Department of Physics, P.O. Box 9,
FIN-00014 University of Helsinki, Finland}

\abstract

A three-dimensional 
effective theory of the finite-temperature SU(5)+adjoint Higgs model 
is constructed 
using the method of dimensional reduction.
The resulting theory can be used for computer simulations of the
GUT phase transition of the early universe. 
In this paper the 
transition is analyzed at two-loop level in perturbation theory.
The structure of the
effective theory does not essentially depend on the matter
contents of the original four-dimensional one. Therefore the
analysis of this paper applies also to more realistic
theories.

\endabstract
\end{frontmatter}

\section{Introduction}
The simplest candidate for a theory unifying all the known gauge 
interactions is the SU(5) model proposed by Georgi and Glashow
 in 1974 \cite{ref:Georgi}. Even though this model is known to conflict
with experiments, it might still give a qualitatively correct picture
of the unification. For example the supersymmetric version of this model
is compatible with the experiments.

Irrespective of the exact form of the unified theory, it is generally
believed that in the early stages of the evolution of our universe,
there was a phase transition in which the unified symmetry broke
down to the residual SU(3)$\times$SU(2)$\times$U(1) symmetry.
The details of this transition may have had important cosmological
consequences. The change in vacuum energy density may have given rise
to an exponential expansion of the universe \cite{ref:inflaatio}.
The recently in \cite{ref:thermalinf} proposed scenario of thermal inflation 
is even more closely connected to the GUT transition.
Due to the non-trivial first homotopy of the residual symmetry group
a large amount of magnetic monopoles may have been created in the 
transition, having dramatic effects on the evolution of the universe
\cite{ref:Preskill,ref:Kibble}. Recently it has also been proposed
that the monopole problem could be solved by symmetry non-restoration 
of the SU(5) theory at high
temperature \cite{ref:salomonson,ref:dvali,ref:bimonte}.

Various properties of the high-temperature SU(5)
model have been analysed by many authors 
\cite{ref:Kuzmin,ref:cook,ref:billoire,ref:guth}. 
Since the model as such is too complicated for efficient computer simulations,
the studies have been mostly based on perturbation theory. 
It is 
well known
that many of the properties of the phase transitions cannot be
obtained by perturbative calculations, since the massless particles
cause infrared divergences \cite{ref:linde}.
However, the only existing lattice simulations are more than 
a decade old and
are therefore very inaccurate according to present standards 
\cite{ref:Olynyk}.
The method of dimensional reduction 
\cite{ref:Ginsparg,ref:Appelquist,ref:dimred}
gives a way to split the problem into two parts, one of which can be handled
perturbatively and the other with computer simulations. 
The idea is to first 
integrate out the Matsubara modes which at high temperatures
decouple from the zero mode and then integrate out the heavy temporal component
of the gauge field. The resulting 3d theory can then be simulated on
a lattice using the results of \cite{ref:hilapert}. 
This procedure has been successfully applied to the electroweak 
phase transition
\cite{ref:ewpt} as well as to other models with relatively simple gauge
symmetries \cite{ref:simSU2H,ref:simMSSM,ref:simU1,ref:simQCD,ref:simSU2A}. 
The SU(5) gauge group has a much more
complicated structure and has some qualitatively new features
since in the Lie algebra of SU($N$) with $N>3$ there are two different
symmetric tensors of order four while for $N<4$ there is only one.
This makes the study of the model interesting also independently of all
the cosmological applications.

The paper is organized as follows. In Section \ref{sect:Lagr}
we define the model considered. In Sections \ref{sect:SuperHeavy} and
\ref{sect:Heavy} we apply the method of dimensional reduction to
the model. In Sections \ref{sect:Parameters} and \ref{sect:Properties} we 
examine the relations between 4d and 3d parameters and
discuss the properties of
the resulting effective theory. The two appendices discuss 
the group theoretical
factors and the evaluation of the two-loop effective
potential. 

\section{SU(5)+adjoint Higgs}
\label{sect:Lagr}
The original model suggested in \cite{ref:Georgi} consists of two scalar
fields, one in adjoint and one in fundamental representation of the gauge 
group SU(5), and two fermion fields, one of which is in ${\bf 5^*}$ and the
other in ${\bf 10}$ representation. To get a more realistic theory, one could
even add a third scalar field, which would be in ${\bf 45}$ representation
\cite{ref:higgs45}. Also a model with an additional 
SU(5) singlet scalar
field has been suggested as a scenario for inflation of the universe
\cite{ref:shafi}. All these models contain many free parameters, whose
values cannot be fixed by experiments, since their effects are only
visible near the GUT energy scale. Some constraints can be derived
for example from cosmological consequences. However, from the
running of the Standard Model gauge couplings, the value of the
gauge coupling constant at the GUT scale can be seen to be $g\approx 0.39$.
 
For simplicity, we shall consider a model
with an adjoint scalar field only. Therefore the theory is not really
physical and we will not fix the parameter values to any fenomenologically
favourable ones, but we will instead explore the whole space of the
parameters. 
Nevertheless, the calculations can be straightforwardly
extended to include the remaining fields, all of which would be integrated
out during dimensional reduction. Therefore the resulting
effective theory would be of exactly the same form, only the relations
between the 3d and 4d parameters would differ. This also applies to
supersymmetric extensions of SU(5), although in that case a more complicated
effective theory may be needed \cite{ref:bjls}.

Our Minkowskian Lagrangian is
\bea
\label{eq:orig4d}
{\mathcal L}_M&=&-\frac{1}{2}\Tr F_{\mu\nu}F^{\mu\nu}
+\Tr(D_\mu\Phi D^\mu\Phi)
-m^2\Tr\Phi^2
-\lambda_1(\Tr\Phi^2)^2
-\lambda_2\Tr\Phi^4,
\eea
where $A_\mu=A_\mu^A T^A$ is the gauge field and 
$\Phi=\Phi^A T^A$ is the Higgs field. The
generators $T^A$ of the symmetry
group are Hermitian and normalized such that
$
\Tr T^AT^B=\frac{1}{2}\delta^{AB}.
$
Their properties are discussed in more detail in Appendix \ref{app:SU5}.
The covariant derivative is
$
D_\mu\Phi=\partial_\mu\Phi+i g[A_\mu,\Phi].
$
One should note that many authors have chosen a different convention. For
example Langacker \cite{ref:Langacker} uses
$
g_{\rm Langacker}=-\sqrt{2}g.
$

Perturbatively we have in our system 24 massless vector fields and 24
scalar fields in the symmetric phase. In the physical broken phase,
there are 12 massless and 12 massive vectors
and 12 scalar fields.
Eight of the scalars form an octet in SU(3),
three form an triplet in SU(2) and one is charged with respect to U(1).
The massless vectors are the gauge bosons of the residual symmetry.
On tree-level the observable parameters in the broken phase are the masses
of the fields, given in Eqs.~(\ref{equ:vectormass}), (\ref{equ:brokenmass}).
When one takes the loop corrections into account, they are replaced
by the pole masses. These differ from the renormalized tree-level masses,
when $\overline{\rm MS}$ scheme is used. 

Because of confinement, the real spectrum of the theory differs completely
from the perturbative one. All the observable fields are SU(5) singlets
and are therefore composite operators of the perturbative fields. The simplest
gauge
invariant operators are the following:
\beq
\Tr\Phi^n,\quad \Tr F_{\mu\nu}\Phi^{n-1},\quad 2\le n\le 5.
\eeq
In the symmetric phase these operators represent bound
states but in the broken phase they can be identified with the
perturbative fields. All the observables of the system are correlators
of these, or more complicated operators, which can be interpreted
as bound states also in the broken phase.

At a finite temperature $T$ the system can be described with the same
Lagrangian with an Euclidian time dimension in which the
bosonic fields are periodic with period $\frac{1}{T}$.
Let us therefore replace the time variable $t$ with $-i\tau$. 
For the calculations we will adopt the Landau
gauge and hence we write the Euclidian Lagrangian as
\bea
{\mathcal L}_E&=&\frac{1}{2}\Tr F_{\mu\nu}F_{\mu\nu}
+\Tr(D_\mu\Phi)^2
+m^2\Tr\Phi^2
+\lambda_1(\Tr\Phi^2)^2
+\lambda_2\Tr\Phi^4\nn
&&
+\frac{1}{2\xi}(\partial_\mu A_\mu^A)^2
+\partial_\mu \overline{c}^A\partial_\mu c^A
-gf^{ABC}\partial_\mu\overline{c}^A A_\mu^Bc^C,
\eea
where $c$ is a ghost field and $\xi\rightarrow 0$.

\section{Superheavy scale}
\label{sect:SuperHeavy}
The first task is to integrate over the superheavy, non-static 
Matsubara modes. At this
stage all the fermions of the theory would disappear, since they do not
have a static mode owing to their antiperiodicity. The time component $A_0$
of the gauge field will become an adjoint scalar field. We shall
start by writing as an Ansatz the most general Lagrangian with at most
quartic interactions which respects the desired symmetries and contains
two adjoint scalar fields and a gauge field:
\bea
\label{equ:3dlagr}
{\mathcal L}_3&=&\frac{1}{4}F_{ij}^AF_{ij}^A
+\Tr(D_i\Phi)^2
+m_3^2\Tr\Phi^2
+\lambda_1'(\Tr\Phi^2)^2
+\lambda_2'\Tr\Phi^4\nn
&&
+\Tr(D_iA_0)^2
+m_3^2\Tr A_0^2
+\kappa_1(\Tr A_0^2)^2
+\kappa_2\Tr A_0^4\nn
&&
+\alpha_1\Tr\Phi^2\Tr A_0^2
+\alpha_2(\Tr\Phi A_0)^2
+\alpha_3\Tr\Phi^2 A_0^2
+\alpha_4\Tr\Phi A_0\Phi A_0.
\eea
The fields and the coupling constants have dimensions
\beq
[A_\mu]=[\Phi]=[g_3]={\rm GeV}^{1\over 2},\quad
[\lambda_i']=[\kappa_i]=[\alpha_i]=[m_3]={\rm GeV}.
\eeq

A renormalizable theory could have interaction vertices of as many as six
fields but we neglect these higher-order corrections \cite{ref:dimred}.
We shall now extract the parameters of (\ref{equ:3dlagr}) and the correct
normalization of the fields by comparing the two- and four-point Green's
functions of the finite temperature 4d theory and of the 3d theory. 
This has been explained 
in great detail in \cite{ref:dimred}.

\input{kuva1}

For the normalization of the fields the parts proportional to the
momentum squared $k^2$
are needed in two-point correlators
$\langle\Phi\Phi\rangle$, $\langle A_0 A_0\rangle$
and $\langle A_i A_j\rangle$. The corresponding diagrams
are shown in Fig.~\ref{fig:normitus}. We will write down only
the part coming from the non-static modes of the diagrams.
The high temperature
approximation of the non-zero modes gives
\bea
(1.1)&=&\frac{k^2}{16\pi^2}g^2\delta^{AB}\frac{15}{\epsilon_b},\\
(1.2)&=&\frac{k^2}{16\pi^2}g^2\delta^{AB}
({10\over \epsilon_b}-\frac{25}{3}),\\
(1.3)&=&\frac{k^2}{16\pi^2}(\delta_{ij}-\frac{k_ik_j}{k^2})g^2\delta^{AB}
({10\over \epsilon_b}+\frac{5}{3}).
\eea
Here 
\beq
\frac{1}{\epsilon_b}=\frac{1}{\varepsilon}+L_b=\frac{1}{\varepsilon}+
\log\frac{\mu^2}{(4\pi T)^2}+2\gamma_E,
\eeq
where $\gamma_E\approx0.577216$ is the Euler-Mascheroni constant. 

We will also need the Debye mass of the $A_0$-field. This is given by
the constant part of $\langle A_0 A_0\rangle$. The corresponding diagrams 
are also given in Fig.~\ref{fig:normitus}. We get
\bea
(1.4)&=&-g^2\delta^{AB}(\frac{5}{2}T^2+\frac{m^2}{16\pi^2}10).
\eea

The effective mass of the Higgs field $\Phi$ could also be evaluated
in a similar way, but we need it to two-loop order and it is easier
to calculate it using the effective potential. We shall postpone this, 
since in the calculation the counterterms from the four-point
correlators are needed. The gauge field $A_i$ does not acquire a mass
in dimensional reduction due to gauge invariance.

\input{kuva2}

The three-dimensional gauge coupling constant $g_3$ is most easily
obtained from the $\langle\Phi\Phi A_iA_j\rangle$-correlator. This consists
of the diagrams shown in Fig.~\ref{fig:4point}. The result is
\bea
(2.1)&=&\frac{g^4}{16\pi^2}\delta_{ij}
(f^{ACE}f^{BDE}+f^{ADE}f^{BCE})\frac{15}{2}\frac{1}{\epsilon_b}.
\eea
Exactly the same diagrams give also the $\langle\Phi\Phi A_0A_0\rangle$
-correlator:
\bea
(2.2)&=&\frac{g^2}{16\pi^2}
\left[
-(12g^2+20\lambda_1)\delta^{AB}\delta^{CD}
-(6g^2+2\lambda_2)(\delta^{AD}\delta^{BC}+\delta^{AC}\delta^{BD})
\right.\nn
&&\left.
-(15g^2-10\lambda_2)d^{ABE}d^{CDE}
\right.\nn
&&\left.
+\left(
\frac{15}{2}\frac{1}{\epsilon_b}g^2-5g^2-4\lambda_1-\frac{8}{5}\lambda_2
\right)
(f^{ACE}f^{BDE}+f^{ADE}f^{BCE})\right].
\eea
We need also the self-coupling of $A_0$, for which one gets
\bea
(2.3)&=&
\frac{g^4}{16\pi^2}
\left[
(\delta^{AB}\delta^{CD}+\mbox{perm.})
+\frac{5}{8}
(d^{ABE}d^{CDE}+\mbox{perm.})\right]
\left(-16\right).
\eea
The notation ``perm.'' denotes all the possible different permutations
of indices, that is
\bea
\delta^{AB}\delta^{CD}+\mbox{perm.}&=&
\delta^{AB}\delta^{CD}+\delta^{AC}\delta^{BD}+\delta^{AD}\delta^{BC},\nn
d^{ABE}d^{CDE}+\mbox{perm.}&=&
d^{ABE}d^{CDE}+d^{ACE}d^{BDE}+d^{ADE}d^{BDE}.
\eea

The last four-point function to be considered is 
$\langle\Phi\Phi\Phi\Phi\rangle$, which gives
\bea
\label{equ:phi4}
(2.4)&=&
\frac{1}{16\pi^2}
\left\{
\left[
\left(
64\lambda_1^2+\frac{212}{5}\lambda_1\lambda_2+\frac{232}{25}\lambda_2^2
+12g^4\right)(\delta^{AB}\delta^{CD}+\mbox{perm.})
\right.\right.\nn
&&\left.\left.
+\left(
12\lambda_1\lambda_2+\frac{32}{5}\lambda_2^2+\frac{15}{2}g^4
\right)(d^{ABE}d^{CDE}+\mbox{perm.})
\right]
\frac{1}{\epsilon_b}\right.\nn
&&\left.
-g^4[8(\delta^{AB}\delta^{CD}+\mbox{perm.})
+5(d^{ABE}d^{CDE}+\mbox{perm.})]\right\}.
\eea

Let us now calculate the effective mass of the Higgs field using the
effective potential.
For this we must shift the
Higgs field. At this point the direction of the shift is irrelevant,
but we choose the physical one
\beq
\Phi\mapsto\Phi+\frac{v}{\sqrt{15}}
{\rm Diag}(1,1,1,-\frac{3}{2},-\frac{3}{2})=\Phi+v\tau_1.
\eeq

After the shift we have 12 massive vector bosons with mass
\beq
\label{equ:vectormass}
M^2=\frac{5}{12}g^2v^2.
\eeq
There are also four different kinds of Higgs fields, one for each factor
group of the residual group and one Goldstone. They have the masses
\bea
\label{equ:brokenmass}
m_1^2&=&m^2+\left(\lambda_1+\frac{2}{5}\lambda_2\right)v^2,\quad
m_2^2=m^2+\left(\lambda_1+\frac{7}{30}\lambda_2\right)v^2,\nn
m_3^2&=&m^2+\left(\lambda_1+\frac{9}{10}\lambda_2\right)v^2,\quad
m_4^2=m^2+\left(3\lambda_1+\frac{7}{10}\lambda_2\right)v^2.
\eea

We will also need explicit expressions for the renormalization
counterterms in the broken phase to one-loop order. 
The mass counterterms in the $\overline{\mbox{\rm MS}}$-scheme are
\bea
\delta m_1^2&=&
  \frac{1}{16\pi^2}\frac{1}{\varepsilon}
\left[m^2 \left(26 \lambda_1+\frac{47}{5} \lambda_2\right)
\right.\nn
&&\left.
+
               v^2 \left(32 \lambda_1^2 + 
\frac{118}{5} \lambda_1 \lambda_2 + \frac{148}{25} \lambda_2^2 
+ \frac{15}{2} g^4\right)\right]  ,\\
\delta m_2^2&=&
  \frac{1}{16\pi^2}\frac{1}{\varepsilon}
  \left[m^2 \left(26 \lambda_1+\frac{47}{5} \lambda_2\right)
\right.\nn
&&\left.
+
               v^2\left(32 \lambda_1^2 + \frac{108}{5} \lambda_1 \lambda_2 
+ \frac{364}{75} \lambda_2^2 + \frac{25}{4} g^4\right)\right],\\
\delta m_3^2&=&
  \frac{1}{16\pi^2}\frac{1}{\varepsilon}
  \left[m^2 \left(26 \lambda_1+\frac{47}{5} \lambda_2\right)
\right.\nn
&&\left.
+
               v^2\left(32 \lambda_1^2 + \frac{148}{5} \lambda_1 \lambda_2 
+ \frac{228}{25} \lambda_2^2 + \frac{45}{4} g^4\right)\right],\\ 
\delta m_4^2&=&
  \frac{1}{16\pi^2}\frac{1}{\varepsilon}
  \left[m^2\left(26 \lambda_1+\frac{47}{5} \lambda_2\right)
\right.\nn
&&\left.
+
               v^2\left(96 \lambda_1^2 + \frac{324}{5} \lambda_1 \lambda_2 
+ \frac{364}{25} \lambda_2^2 + \frac{75}{4} g^4\right)\right],\\
\delta M^2&=&
  \frac{1}{16\pi^2}\frac{1}{\varepsilon}
\frac{25}{8}g^4v^2,
\eea
and the wave function counterterms are
\bea
\delta Z_\Phi&=&  \frac{g^2}{16\pi^2}\frac{1}{\varepsilon}15,\\
\delta Z_A&=&  \frac{g^2}{16\pi^2}\frac{1}{\varepsilon}10.
\eea

\input{kuva3}

The one-particle irreducible vacuum diagrams needed are shown in 
Fig.~\ref{fig:Efpot}. We need only the part of the result which is
quadratic in $v$. 
We have written down also the terms quartic in $v$ to one-loop order just
as a check of the result (\ref{equ:phi4}). Using the color factors given in
Appendix \ref{app:2loopep} we could make the full two-loop calculation,
but it is not necessary here.
In the high temperature approximation the result
is
\bea
V(v)&=&V^{3d}(v)+\mbox{constant}\nn
&&+\frac{1}{2}v^2\left\{
\left(
\frac{5}{4}g^2+\frac{13}{6}\lambda_1+\frac{47}{60}\lambda_2
\right)T^2\right.\nn
&&\left.
-
\frac{m^2}{16\pi^2}
\left(26\lambda_1+\frac{47}{5}\lambda_2\right)L_b\right.\nn
&&\left.
+\frac{T^2}{16\pi^2}\left[
\left(\frac{1}{\varepsilon}-4\log\frac{3T}{\mu}-4c\right)
\left(
13\lambda_1^2+\frac{47}{5}\lambda_1\lambda_2\right.\right.\right.\nn
&&\left.\left.\left.+\frac{493}{100}\lambda_2^2
-65g^2\lambda_1-\frac{47}{2}g^2\lambda_2\right)\right.\right.\nn
&&\left.\left.
+L_b
\left(-\frac{25}{2}g^4+\frac{65}{2}g^2\lambda_1+\frac{47}{4}g^2\lambda_2
-\frac{208}{3}\lambda_1^2-\frac{752}{15}\lambda_1\lambda_2
-\frac{922}{75}\lambda_2^2\right)
\right.\right.\nn
&&\left.\left.
+\frac{275}{12}g^4+\frac{65}{3}g^2\lambda_1+\frac{47}{6}g^2\lambda_2\right]
\right\}\nn
&&+\frac{1}{24}v^4\frac{1}{16\pi^2}
\left[
\left(25-\frac{75}{2}L_b\right)g^4\right.\nn
&&\left.
-\left(192\lambda_1^2+\frac{648}{5}\lambda_1\lambda_2
+\frac{728}{25}\lambda_2^2\right)L_b\right],
\eea
where $c=\frac{1}{2}\left(\log\frac{8\pi}{9}+\frac{\zeta'(2)}{\zeta(2)}
-2\gamma_E\right).$

Now we have all the necessary Green's functions to fix the three-dimensional
parameters. This is done by matching the corresponding results
in both theories taking into account that the normalization of the fields
is also different. At one-loop level one only has to add
corrections to the coupling constants and the field renormalization.
The relations between the three-dimensional and four-dimensional
fields are
\bea
\Phi_{3d}^2&=&\frac{1}{T}
\left[1-\frac{g^2}{16\pi^2}15L_b\right]\Phi^2,\\
(A_0^{3d})^2&=&\frac{1}{T}
\left[1-\frac{g^2}{16\pi^2}\left(10L_b-\frac{25}{3}\right)\right]A_0^2,\\
(A_i^{3d})^2&=&\frac{1}{T}
\left[1-\frac{g^2}{16\pi^2}\left(10L_b+\frac{5}{3}\right)\right]A_i^2.
\eea

The parameters of the three-dimensional theory written as functions of
those of the original 4d theory are
\bea
g_3^2&=&g^2T\left[1+\frac{g^2}{16\pi^2}\left(\frac{35}{2}L_b+\frac{5}{3}\right)
\right],\\
\kappa_1&=&\frac{g^4T}{16\pi^2}6,\\
\kappa_2&=&\frac{g^4T}{16\pi^2}10,\\
\lambda_1'&=&\lambda_1T
-\frac{T}{16\pi^2}
\left[
\left(
32\lambda_1^2+\frac{94}{5}\lambda_1\lambda_2+\frac{84}{25}\lambda_2^2
+\frac{9}{2}g^4-30g^2\lambda_1\right)L_b\right.\nn
&&\left.-3g^4\right],\\
\lambda_2'&=&\lambda_2T
-\frac{T}{16\pi^2}
\left[
\left(
12\lambda_1\lambda_2+\frac{32}{5}\lambda_2^2
+\frac{15}{2}g^4-30g^2\lambda_2\right)L_b-5g^4\right],\\
\label{equ:alpha1}
\alpha_1&=&\frac{g^2T}{16\pi^2}\left(6g^2+24\lambda_1\right),\\
\alpha_2&=&\frac{g^2T}{16\pi^2}\left(12g^2+4\lambda_2\right),\\
\alpha_3&=&2g^2T\left[
1+\frac{1}{16\pi^2}\left(
\frac{35}{2}L_bg^2+\frac{35}{3}g^2+4\lambda_1-\frac{42}{5}\lambda_2\right)
\right],\\
\label{equ:alpha4}
\alpha_4&=&-2g^2T\left[
1+\frac{1}{16\pi^2}\left(
\frac{35}{2}L_bg^2-\frac{10}{3}g^2+4\lambda_1+\frac{8}{5}\lambda_2\right)
\right],\\
m_D^2&=&\frac{5}{2}g^2T^2,\\
\label{equ:m3}
m_3^2&=&\tilde{m}^2+
\frac{T}{12}\left(15g_3^2+26\lambda_1'+\frac{47}{5}\lambda_2'\right)\nn
&&
+\frac{1}{16\pi^2}\frac{5}{6}g_3^2\left(
\frac{25}{2}g_3^2+26\lambda_1'+\frac{47}{5}\lambda_2'\right)\nn
&&+\frac{1}{16\pi^2}\left[
10g_3^2\left(26\lambda_1'+\frac{47}{5}\lambda_2'\right)
-52\lambda_1'^2\right.\nn
&&\left.-\frac{188}{5}\lambda_1'\lambda_2'
-\frac{493}{25}\lambda_2'^2\right]
\left(\log\frac{3T}{\mu}+c\right),\\
\tilde{m}^2&=&m^2\left[
1+\frac{1}{16\pi^2}\left(15g^2-26\lambda_1-\frac{47}{5}\lambda_2\right)L_b\right].
\eea

In every equation the $\mu$-dependence of $L_b$ cancels the $\mu$-dependence
of the coupling constants. Therefore none of the coupling constants of the
three-dimensional theory runs.
From Eq.~(\ref{equ:m3}) one can read the running of the 
three-dimensional mass as a function of the scale $\mu$,
\bea
\label{equ:f_2m}
\mu\frac{\partial m_3^2(\mu)}{\partial \mu}&=&-\frac{1}{16\pi^2}f_{2m}\nn
&=&
-\frac{1}{16\pi^2}
\left[
10g_3^2\left(26\lambda_1'+\frac{47}{5}\lambda_2'\right)
-52\lambda_1'^2-\frac{188}{5}\lambda_1'\lambda_2'
-\frac{493}{25}\lambda_2'^2\right].
\eea
As we can see later in Eq.~(\ref{equ:finalmass}), the running due to
the coupling of the two adjoint scalar fields $\Phi$ and $A_0$ cancels
when the coupling constants $\alpha$ have the values given in 
Eqs.~(\ref{equ:alpha1})--(\ref{equ:alpha4}). Therefore the $A_0$ field
does not contribute to $f_{2m}$ and the result (\ref{equ:f_2m}) 
is the same
also for a theory with only one adjoint scalar field.
This result can be generalized to the case of SU($N$) symmetry,  and is
\bea
f_{2m}&=&
2Ng_3^2\left((N^2+1)\lambda_1'+(2N^2-3)\frac{\lambda_2'}{N}\right)
\nn
&&
-2(N^2+1)\lambda_1'^2-4(2N^2-3)\lambda_1'\frac{\lambda_2'}{N}
-\left(N^4-6N^2+18\right)\frac{\lambda_2'^2}{N^2}.
\eea
Since the three-dimensional theory is superrenormalizable, this result
is exact for a theory with only one adjoint Higgs.

Thus we have now constructed a three-dimensional theory with a gauge field
and two adjoint Higgses, which describes the same physics as the original
theory in a sense that the static
correlators of the three-dimensional theory
coincide with those of the four-dimensional one at order ${\mathcal O}(g^4)$.
However, the theory is still unnecessarily complicated. At high
temperature the $A_0$ field is namely heavy, since its mass is
proportional to the temperature. Thus one can integrate it out as well.
We will do that in the next section.

\section{Heavy scale}
\label{sect:Heavy}

\input{kuva4}

Integrating out the heavy $A_0$ field is also described in \cite{ref:dimred}.
For that task we need all the two- and four-point diagrams containing the
$A_0$ field. They are shown in Fig.~\ref{fig:heavy}.

On one-loop level there are no momentum dependent two-point $\langle\Phi\Phi
\rangle$-diagrams with the $A_0$ field. Therefore the normalization of
the Higgs field does not change. For the gauge field $A_i$
we do instead have one diagram:
\bea
(4.1)&=&-\frac{g_3^2}{4\pi m_D}\frac{5}{12}
\delta^{AB}(k^2\delta_{ij}-k_ik_j).
\eea

The correction to the gauge coupling can be most easily evaluated from
the $\langle\Phi\Phi A_iA_i\rangle$-correlator. There are two such 
diagrams
with the $A_0$ field, but they cancel each other:
\bea
(4.2.1)&=& 
\frac{g_3^2\delta_{ij}}{16\pi m_D}
\left[
(10\alpha_1+2\alpha_3)\delta^{AB}\delta^{CD}
-\alpha_4(\delta^{AC}\delta^{BD}+\delta^{AD}\delta^{BC})\right.\nn
&&\left.
+\alpha_2(f^{ACE}f^{BDE}+f^{ADE}f^{BCE})
+\frac{5}{2}\alpha_3d^{ABE}d^{CDE}\right]
,\\
(4.2.2)&=&-(4.2.1),\\
(4.2)&=&0.
\eea
Thus 
the only effect to the
gauge coupling will be from the new normalization of the
gauge field. 

The correction for the Higgs self-interaction can be obtained from
the only $\langle\Phi\Phi\Phi\Phi\rangle$-diagram with $A_0$-lines.
The result is
\bea
(4.3)&=&
\frac{1}{16\pi m_D}
\left[
\left(
24\alpha_1^2+2\alpha_1\alpha_2+\frac{48}{5}\alpha_1\alpha_3
+\frac{2}{5}\alpha_2\alpha_3\right.\right.\nn
&&\left.\left.+\frac{24}{25}\alpha_3^2
-\frac{2}{5}\alpha_1\alpha_4+\frac{2}{5}\alpha_2\alpha_3
-\frac{2}{25}\alpha_3\alpha_4+\frac{24}{25}\alpha_4^2\right)
(\delta^{AB}\delta^{CD}+\mbox{perm.})\right.\nn
&&\left.
+\left(
\alpha_2\alpha_3+\frac{21}{20}\alpha_3^2+\alpha_2\alpha_4
-\frac{2}{5}\alpha_3\alpha_4-\frac{1}{5}\alpha_4^2\right)
(d^{ABE}d^{CDE}+\mbox{perm.})\right].\nn
\eea

To get the mass correction of the Higgs field we shall evaluate the 
$A_0$-dependent part of the two-loop
effective potential. Breaking the symmetry gives for the
gauge field $A_i$ and the Higgs field $\Phi$ the same masses as before
in Eqs.~(\ref{equ:vectormass}), (\ref{equ:brokenmass}). 
The $A_0$ field will
acquire the masses
\bea
M_1^2&=&m_D^2
+(\frac{1}{2}\alpha_1+\frac{1}{15}\alpha_3+\frac{1}{15}\alpha_4)v^2,\\
M_2^2&=&m_D^2
+(\frac{1}{2}\alpha_1+\frac{13}{120}\alpha_3-\frac{1}{10}\alpha_4)v^2,\\
M_3^2&=&m_D^2
+(\frac{1}{2}\alpha_1+\frac{3}{20}\alpha_3+\frac{3}{20}\alpha_4)v^2,\\
M_4^2&=&m_D^2
+(\frac{1}{2}\alpha_1+
\frac{1}{2}\alpha_2+\frac{7}{60}\alpha_3+\frac{7}{60}\alpha_4)v^2.
\eea
Now the desired part of the effective potential can be obtained from
the six vacuum diagrams shown in Fig.~\ref{fig:heavy}.
They give
\bea
(4.4.1)&=&  8 C_{\rm S}(M_1)+12 C_{\rm S}(M_2)
+3 C_{\rm S}(M_3)+C_{\rm S}(M_4),\\
(4.4.2)&=&
-  (20 \kappa_1+10 \kappa_2) D_{\rm SS}(M_1,M_1)-
   (42 \kappa_1+15 \kappa_2) D_{\rm SS}(M_2,M_2)\nn
&&-
   \left(\frac{15}{4} \kappa_1+\frac{15}{8} \kappa_2\right) D_{\rm SS}(M_3,M_3)
   -\left(\frac{3}{4} \kappa_1+\frac{7}{40} \kappa_2\right) D_{\rm SS}(M_4,M_4)\nn
&&-
   (48 \kappa_1+16 \kappa_2) D_{\rm SS}(M_1,M_2)-
   12 \kappa_1 D_{\rm SS}(M_1,M_3)\nn
&&
   -\left(4 \kappa_1+\frac{8}{5} \kappa_2\right) D_{\rm SS}(M_1,M_4)-
   (18 \kappa_1+9 \kappa_2) D_{\rm SS}(M_2,M_3)\nn
&&-
   \left(6 \kappa_1+\frac{7}{5} \kappa_2\right) D_{\rm SS}(M_2,M_4)-
   \left(\frac{3}{2} \kappa_1+\frac{27}{20} \kappa_2\right) D_{\rm SS}(M_3,M_4),\\
(4.4.3)&=&
-\left(16\alpha_1+2\alpha_2+\frac{16}{3}\alpha_3-\frac{2}{3}\alpha_4\right)
D_{\rm SS}(m_1,M_1)\nn
&&
-\left(36\alpha_1+3\alpha_2+\frac{15}{2}\alpha_3 \right)
D_{\rm SS}(m_2,M_2)\nn
&&
-\left(\frac{9}{4}\alpha_1+\frac{3}{4}\alpha_2
+\frac{9}{8}\alpha_3-\frac{3}{8}\alpha_4\right)
D_{\rm SS}(m_3,M_3)\nn
&&
-\left(\frac{1}{4}\alpha_1+\frac{1}{4}\alpha_2
+\frac{7}{120}\alpha_3+\frac{7}{120}\alpha_4\right)
D_{\rm SS}(m_4,M_4)\nn
&&
-\left(24\alpha_1+4\alpha_3\right)
(D_{\rm SS}(m_1,M_2)+D_{\rm SS}(m_2,M_1))\nn
&&
-6\alpha_1
(D_{\rm SS}(m_1,M_3)+D_{\rm SS}(m_3,M_1))\nn
&&
-\left(2\alpha_1+\frac{4}{15}\alpha_3
+\frac{4}{15}\alpha_4\right)
(D_{\rm SS}(m_1,M_4)+D_{\rm SS}(m_4,M_1))\nn
&&
-\left(9\alpha_1+\frac{9}{4}\alpha_3\right)
(D_{\rm SS}(m_2,M_3)+D_{\rm SS}(m_3,M_2))\nn
&&
-\left(3\alpha_1+\frac{13}{20}\alpha_3
-\frac{3}{5}\alpha_4\right)
(D_{\rm SS}(m_2,M_4)+D_{\rm SS}(m_4,M_2))\nn
&&
-\left(\frac{3}{4}\alpha_1
+\frac{9}{40}\alpha_3+\frac{9}{40}\alpha_4\right)
(D_{\rm SS}(m_3,M_4)+D_{\rm SS}(m_4,M_3)),\nn
(4.4.4)&=&
-\frac{g_3^2}{4} [24 D_{\rm SV}(M_1,0)+30 D_{\rm SV}(M_2,0)
+6 D_{\rm SV}(M_3,0)\nn
&&+16 D_{\rm SV}(M_1,M)+
           30 D_{\rm SV}(M_2,M)+9 D_{\rm SV}(M_3,M)
\nn
&&+5 D_{\rm SV}(M_4,M))],\\
(4.4.5)&=&
-\frac{g_3^2}{4} [24 D_{\rm SSV}(M_1,M_1,0)+30 D_{\rm SSV}(M_2,M_2,0)\nn
&&+6 D_{\rm SSV}(M_3,M_3,0)+
           32 D_{\rm SSV}(M_1,M_2,M)\nn
&&+18 D_{\rm SSV}(M_2,M_3,M)+
           10 D_{\rm SSV}(M_2,M_4,M)]\\
(4.4.6)&=&
-\frac{1}{4}v^2
\left[
24 \alpha_1^2+2 \alpha_1 \alpha_2+\frac{25}{2} \alpha_2^2
+\frac{48}{5} \alpha_1 \alpha_3+\frac{23}{5} \alpha_2 \alpha_3
+\frac{537}{100} \alpha_3^2\right.\nn
&&\left.-\frac{2}{5} \alpha_1 \alpha_4
+\frac{48}{5} \alpha_2 \alpha_4-\frac{44}{25} \alpha_3 \alpha_4
+\frac{581}{50} \alpha_4^2
\right] D_{\rm SSS}(m,m_D,m_D).
\eea
The explicit expressions of the functions $C$ and $D$ are given in
Appendix \ref{app:2loopep}.

Using these results we can construct an effective 
three-dimensional theory with
only one scalar field and the gauge field. The Lagrangian of the
theory is
\bea
\label{equ:effth}
{\mathcal L}&=&\frac{1}{4}F_{ij}^AF_{ij}^A
+\Tr(D_i\Phi)^2
+\overline{m}^2\Tr\Phi^2
+\overline\lambda_1(\Tr\Phi^2)^2
+\overline\lambda_2\Tr\Phi^4,
\eea
where the fields are related to the original fields as follows:
\bea
\overline\Phi^2&=&\Phi_{3d}^2\nn
&=&\frac{1}{T}
\left[1-\frac{g^2}{16\pi^2}15L_b\right]\Phi^2,\\
\overline A_i^2&=&\left[1+\frac{5}{48\pi}\frac{g_3^2}{m_D}\right](A_i^{3d})^2
\nn
&=&\frac{1}{T}
\left[1
+\frac{g}{24\pi}\sqrt{\frac{5}{2}}
-\frac{g^2}{16\pi^2}\left(10L_b+\frac{5}{3}\right)\right]A_i^2.
\eea

The relation of the parameters of the theory to those of the original
theory is
\bea
\label{equ:gviiva}
\overline g^2&=&g_3^2\left[1-\frac{5}{48\pi}\frac{g_3^2}{m_D}\right]\nn
&=&
g^2T\left[
1-\frac{g}{24\pi}\sqrt{\frac{5}{2}}+\frac{g^2}{16\pi^2}
\left(\frac{35}{2}L_b+\frac{5}{3}\right)\right],\\
\label{equ:l1viiva}
\overline\lambda_1&=&
\lambda_1'-\frac{1}{16\pi m_D}\left(
12\alpha_1^2+\alpha_1\alpha_2+\frac{1}{2}\alpha_2^2
+\frac{24}{5}\alpha_1\alpha_3\right.\nn
&&\left.+\frac{27}{100}\alpha_3^2
-\frac{1}{5}\alpha_1\alpha_4+\frac{1}{25}\alpha_3\alpha_4+\frac{13}{25}
\alpha_4^2\right)\nn
&=&
T\left\{\lambda_1-\frac{3g^3}{8\sqrt{10}\pi}\right.\nn
&&\left.
-\frac{1}{16\pi^2}
\left[
\left(
32\lambda_1^2+\frac{94}{5}\lambda_1\lambda_2+\frac{84}{25}\lambda_2^2
+\frac{9}{2}g^4-30g^2\lambda_1\right)L_b-3g^4\right]\right\},\\
\label{equ:l2viiva}
\overline\lambda_2&=&
\lambda_2'-\frac{1}{16\pi m_D}\left(
\alpha_2\alpha_3+\frac{21}{20}\alpha_3^2+\alpha_2\alpha_4
-\frac{2}{5}\alpha_3\alpha_4-\frac{1}{5}
\alpha_4^2\right)\nn
&=&T\left\{\lambda_2-\frac{g^3}{8\pi}\sqrt\frac{5}{2}\right.\nn
&&\left.
-\frac{1}{16\pi^2}
\left[
\left(
12\lambda_1\lambda_2+\frac{32}{5}\lambda_2^2
+\frac{15}{2}g^4-30g^2\lambda_2\right)L_b-5g^4\right]\right\},\\
\label{equ:finalmass}
\overline m^2&=&
m_3^2-\frac{m_D}{4\pi}\left(
12\alpha_1+\frac{1}{2}\alpha_2+\frac{12}{5}\alpha_3-\frac{1}{10}\alpha_4
\right)\nn
&&+\frac{1}{16\pi^2}\nn
&&\phantom{m}
\left[
\left(\frac{1}{\varepsilon}+1+4\log\frac{\mu}{2m_D}\right)
g_3^2
\left(30\alpha_1+\frac{5}{4}\alpha_2+6\alpha_3-\frac{1}{4}\alpha_4
-\frac{25}{8}g_3^2\right)\right.\nn
&&\phantom{m}
\left.
-\left(\frac{1}{\varepsilon}+2+4\log\frac{\mu}{m_3+2m_D}\right)
\left(
3\alpha_1^2+\frac{1}{4}\alpha_1\alpha_2+\frac{25}{16}\alpha_2^2
+\frac{6}{5}\alpha_1\alpha_3
\right.\right.\nn
&&\phantom{m}
\left.\left.
+\frac{23}{40}\alpha_2\alpha_3
+\frac{537}{800}\alpha_3^2-\frac{1}{20}\alpha_1\alpha_4
+\frac{6}{5}\alpha_2\alpha_4-\frac{11}{50}\alpha_3\alpha_4
+\frac{581}{400}\alpha_4^2\right)\right.\nn
&&\phantom{m}
\left.
+\frac{25}{8}g_3^4
+\left(13\kappa_1+\frac{47}{10}\kappa_2\right)
\left(12\alpha_1+\frac{1}{2}\alpha_2+\frac{12}{5}\alpha_3-\frac{1}{10}\alpha_4
\right)\right]\nn
&=&m_3^2-\frac{5g^3T^2}{4\pi}\sqrt\frac{5}{2}
+\frac{g^4T^2}{16\pi^2}
\left(
\frac{75}{2}\log\frac{m_3+2m_D}{2m_D}-\frac{25}{4}\right).
\eea
One should note that for the correct values of the parameters $\alpha$ 
(\ref{equ:alpha1})--(\ref{equ:alpha4}) the
expression (\ref{equ:finalmass}) is not divergent since the divergences
cancel each other. Therefore the new effective mass $\overline{m}^2$
runs in the same way as $m_3^2$ as a function of the scale $\mu$.
In particular, Eq.~(\ref{equ:f_2m}) is correct also in this case.

\section{Parameters of the effective theory}
\label{sect:Parameters}
Let us now study concretely the values of the parameters of the effective
theory (\ref{equ:effth}) and their relation to the parameters of the
original 4d theory (\ref{eq:orig4d}).
As discussed in Sect.~\ref{sect:Lagr}, it is reasonable to fix only
the gauge coupling constant, and keep the other parameters free. Thus we will 
only assume that the parameters of the theory are such that
the construction of the effective
theory is legitimate.

The effective theory has four parameters: $\overline{g}^2$, 
$\overline{\lambda}_1$, $\overline{\lambda}_2$ and $\overline{m}^2$.
However, these are all dimensionful quantities and we can choose one
to fix the scale and express the theory in terms of three
dimensionless parameters
\bea
y&=&\frac{\overline{m}^2(\overline{g}^2)}{\overline{g}^4},\quad
x_1=\frac{\overline{\lambda}_1}{\overline{g}^2},\quad
x_2=\frac{\overline{\lambda}_2}{\overline{g}^2}.
\eea
Here $y$ has been defined in the $\overline{\mbox{\rm MS}}$-scheme 
with the renormalization scale 
$\mu=\overline{g}^2$.
When the 4d parameters have been renormalized at the scale 
$\mu=4\pi Te^{-\gamma_E}$ and $g=0.39$, we can use 
Eqs. (\ref{equ:vectormass}), 
(\ref{equ:brokenmass}) to replace the 4d parameters with
tree-level masses in Eqs.~(\ref{equ:gviiva})--(\ref{equ:finalmass}) 
and we obtain
\bea
\label{equ:x1x2y}
x_1&=&0.21\frac{m_4^2}{M^2}-0.59\frac{m_1^2}{M^2}-0.012,\quad
x_2=2.52\frac{m_1^2}{M^2}-0.020,\nn
y&=&6.80+6.08\frac{m_1^2}{M^2}-1.45\frac{m_1^4}{M^4}
+3.95\frac{m_4^2}{M^2}-0.038\frac{m_4^4}{M^4}\nn
&&-0.12\frac{m_1^2m_4^2}{M^4}
-21.9\frac{m_4^2}{T^2},
\eea
where we have neglected the logarithmic dependence on the temperature.
Here $M$, $m_1$ and $m_4$ are the tree-level broken phase masses of the massive
vectors, the SU(3) octet scalars and the U(1) charged scalar, respectively.
As discussed in Section \ref{sect:Lagr}, the physical pole masses differ
from these due to loop corrections. 
From Eq.~(\ref{equ:x1x2y}) we see that as the temperature decreases, 
$x_1$ and $x_2$ stay constant
and $y$ decreases. Since the transition takes place near $y=0$, we can
get a simple approximation for the critical temperature by setting 
$y(T=T_c)=0$ in Eq.~(\ref{equ:x1x2y}). In the next section we derive
more accurate estimates.

\section{Properties of the effective theory}
\label{sect:Properties}
In a system with an SU($N$) gauge field coupled to a scalar field in
fundamental representation, it has been shown that the ``symmetric'' and
the ``broken'' phase are analytically connected to each other 
\cite{ref:fradkin}. This has been confirmed by numerical simulations 
in the case $N=2$ \cite{ref:notrans}. 
In \cite{ref:fradkin} the authors
expect that for SU($N$)+adjoint Higgs there could be a symmetry breaking
related to the $Z_N$ symmetry in which the Higgs field transforms 
trivially. 
Nevertheless, lattice results have given
evidence that for SU(2)+adjoint Higgs there is no distinction
between the phases and for some values of the parameters the
phase transition does not exist \cite{ref:simSU2A}. 
It is therefore uncertain, to what extent we can consider
the phase transition to be a symmetry breakdown as suggested by
the perturbative interpretation.

Even though it is well known that only
non-perturbative, numerical computations can give reliable results in the
vicinity of a phase transition, we will inspect the perturbative 
effective potential in order to obtain information about the 
phase diagram.
For small values of $x_1$ and $x_2$ this gives the
correct result. 

The essential shape of the potential is already found in the one-loop
approximation, taking into account only the gauge field loop.
This gives the effective potential
\beq
\label{equ:1loopep}
V(\Phi)=y\Tr\Phi^2+x_1(\Tr\Phi^2)^2+x_2\Tr\Phi^4-\frac{1}{6\pi}
\sum_{i,j=1}^5|\Phi_i-\Phi_j|^3,
\eeq
where $\Phi={\rm Diag}(\Phi_1,\ldots,\Phi_5)$ and $\sum_i\Phi_i=0$
and we have scaled $\Phi$ and $V(\Phi)$ to dimensionless quantities.
The Higgs vev can always be transformed to this diagonal form.
Here $y$ denotes actually the value at the scale $\mu$, 
\beq
y(\mu)=y-\frac{f_{2m}}{16\pi^2}\log\frac{\mu}{\overline{g}^2}.
\eeq
In this approximation the scale $\mu$ cannot be fixed.

In perturbation theory the transition is of first order, due 
to the
cubic term in (\ref{equ:1loopep}). We can evaluate the critical surface
$y_c=y_c(x_1,x_2)$, where the minima of $V(\Phi)$ are degenerate.
In our present approximation this can be calculated analytically.
Let us write $\Phi=v\tau$, where $v$ is a real number and 
$\Tr\tau^2=\frac{1}{2}$.
Then we obtain
\beq
\label{equ:yc1appro}
y_c=\frac{1}{18\pi^2}\frac{(\sum_{i,j}|\tau_i-\tau_j|^3)^2}
{x_1+4x_2\Tr\tau^4}.
\eeq
This result depends of the direction $\tau$ of the symmetry breakdown.
We shall discuss mainly the physically most interesting directions
\beq
\tau_1=\frac{1}{\sqrt{15}}{\rm Diag}(1,1,1,-\frac{3}{2},-\frac{3}{2}),\quad
\tau_2=\frac{1}{2\sqrt{10}}{\rm Diag}(1,1,1,1,-4),
\eeq
which lead to the residual symmetries of the Standard Model and
SU(4)$\times$U(1), respectively.
It has been shown \cite{ref:2omin} that on tree-level these are the only
possible minima. Quantum corrections may change the situation
and for some values of parameters there may be other minima, even a global 
one. 
The corresponding critical surfaces are
\beq
\label{equ:critsurf}
y_c(\tau_1)=\frac{625}{36\pi^2}\frac{1}{30x_1+7x_2},\quad
y_c(\tau_2)=\frac{625}{36\pi^2}\frac{1}{20x_1+13x_2}.
\eeq

From this we see that the critical surfaces cross at the line 
$x_1=\frac{3}{5}x_2$. This ratio is connected to the $Z_5$ symmetry of
the SU(5) gauge theory at finite temperature \cite{ref:ZN}. 
We can namely proceed with the
dimensional reduction for a pure SU(5) theory using essentially the
diagrams already calculated and end up with the ratio $x_1=\frac{3}{5}x_2$
for the effective self-coupling of the $A_0$ field. The origin and the
minima corresponding to the directions $\tau_1$ and $\tau_2$ are
then identified by the $Z_5$ symmetry, which shows up in our
calculation just as the degeneracy in this approximation.
For the other possible minima SU(3)$\times[$U(1)$]^2$
and [SU(2)$]^2\times$[U(1)]$^2$,
the critical surfaces lie lower than the ones shown above for all the
parameter values. 

In addition to the critical surface between the symmetric
phase and the broken phases we are also interested in phase transitions
or critical surfaces
between different broken phases. It turns out that $\tau_1$ and $\tau_2$
are the only shifts that can give an absolute minimum for any values of
the parameters. Therefore only the critical surface between those two phases
is of interest. It consists of the points in the parameter space in which
the minima in the directions $\tau_1$ and $\tau_2$ are degenerate. 
Unfortunately we cannot write a simple expression for that surface
as in Eq.~(\ref{equ:critsurf}).

One can easily improve the accuracy of this one-loop 
approximation. The effective
potential can be calculated to two-loop order using essentially
the color factors calculated previously. However, for some
diagrams this must be done by explicitly fixing the direction
of the shift. Therefore the result cannot be written in a general form
as in (\ref{equ:1loopep}). We have completed this calculation for
the shift directions $\tau_1$ and $\tau_2$ and the expressions for the
effective potential are shown in Appendix \ref{app:2loopep}.
In principle one could write down the complete one-loop effective potential
for a shift in an arbitrary direction, but the resulting expression is
much too complicated and we have omitted it.

\begin{figure}
\centering{\epsfig{file=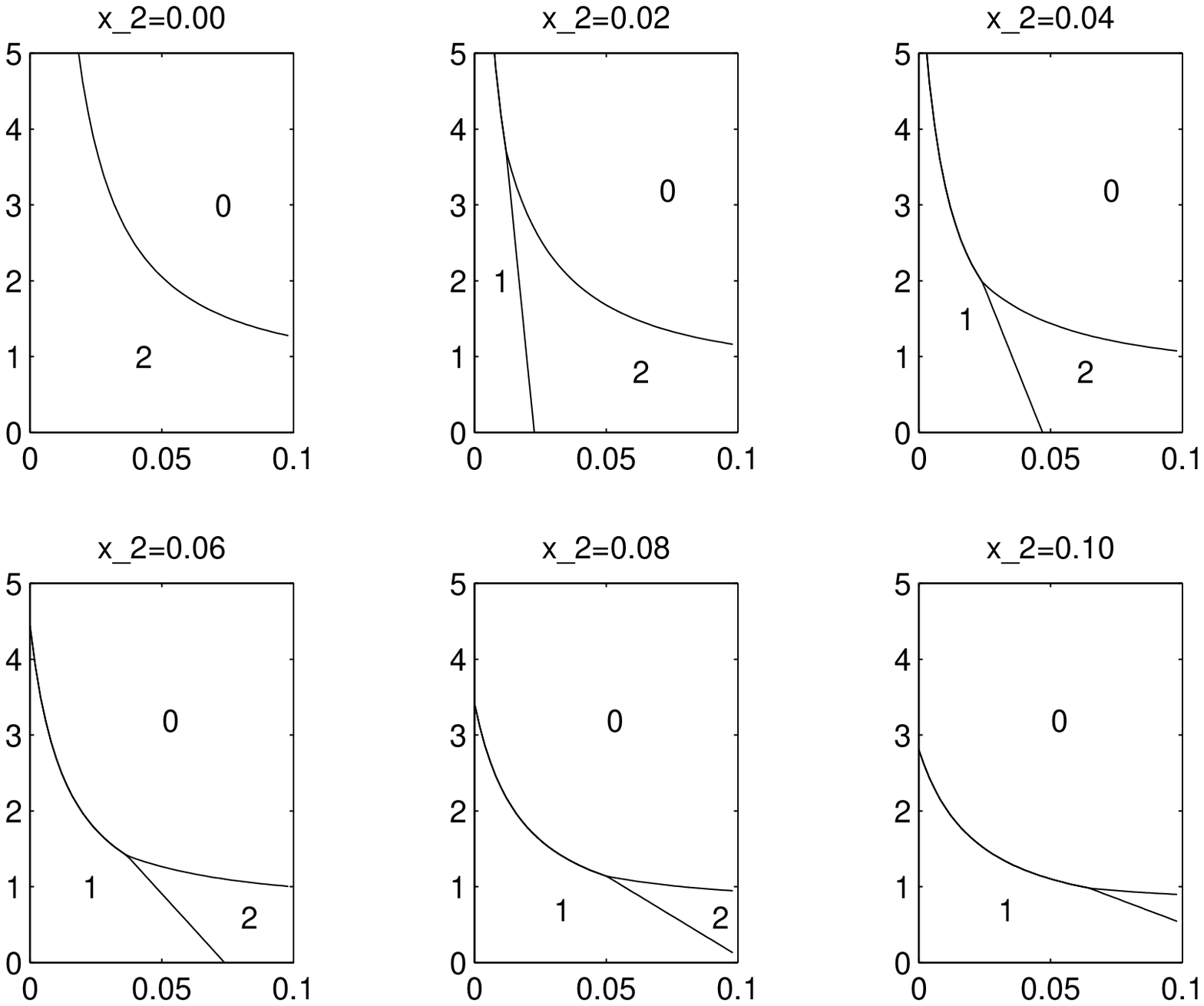,height=10cm,width=13cm}}
\caption{The two-loop phase diagram of the 3d SU(5)+adjoint Higgs model. The 
horizontal and vertical axes correspond to the parameters $x_1$ and $y$,
respectively. The symmetric phase is denoted by 0 and the phases broken to
the directions $\tau_1$ and $\tau_2$ by 1 and 2, respectively.}
\label{fig:critsurf}
\end{figure}

The phase diagram of the system in two-loop approximation is shown in
Fig.~\ref{fig:critsurf}. We have only evaluated the potential
in two broken minima and therefore the possible presence of any
other phases is not shown. The phase diagram consists of three domains
in which the true minimum is in the corresponding phase. The boundaries
between the domains correspond to phase transitions.
This result is correct only
for small values of $x_1$ and $x_2$, but the theory can be simulated
on a lattice to obtain the correct critical surface.
It is interesting to speculate whether the phase transition
becomes a crossover as $x_1$ and $x_2$ grow and become about $0.1$. 
This effect has been found for an SU(2) gauge field coupled to a Higgs
field in the fundamental 
\cite{ref:notrans} or adjoint \cite{ref:simSU2A} representation.
If the same is true for SU(5), the GUT phase transition might never had 
happened.

We cannot get any direct information about the kinetics of the
phase transition from the effective 3d theory, since it only
describes the equlibirium behaviour.
As the parameters cross a critical
surface the two corresponding minima are degenerate, which 
gives rise to phase transition. Owing to the supercooling effects
the transition does not take place instantaneously, but the system may
stay in a metastable state for some time.
A reliable analysis of the kinetics of the phase transition
is possible only after the correct values of $T_c$, latent heat and
interface tension are determined from lattice simulations.

\section{Conclusions}
We have constructed a three-dimensional theory that describes the 
equilibrium behaviour of a high-temperature SU(5)+adjoint Higgs system
correctly up to order ${\mathcal O}(g^3)$. 
This is done by determining the parameter values of the
effective theory by comparing
the Green's functions of the two theories.
The procedure is free from infrared divergences, but solves some of the
problems concerning numerical simulations of the original 4d theory. 
The resulting
theory contains no fermions and has parametrically
only one energy scale. Also the
smaller number of dimensions allows one to use much larger lattice
sizes. Therefore one
can expect to have much more accurate results about the phase transition
than previously. 

We have proceeded with the task of dimensional reduction only for a
non-realistic theory with no fermions. The fermion content of
the theory would only change the values of the parameters of the effective
theory, not the structure of the theory itself, since due to their
antiperiodicity the fermions have only super-heavy Matsubara modes of
mass $\sim \pi T$,
which are integrated out. Near the phase transition the other scalar fields
than the adjoint one are also heavy and therefore do not affect the
structure of the effective theory. Thus we get already now reliable 
qualitative results
from the effective theory. The inclusion of the realistic particle
spectrum is only a mechanical task and involves no new technical
difficulties. One might also extend the analysis to a supersymmetric
version of the theory. Even then an unambiguous picture of the 
phase transition cannot be obtained because of the large freedom in the
choice of the parameters of the four-dimensional theory.

There are a few problems with the construction presented in this paper.
First of all it is questionable how well one can apply the standard
high-temperature formalism in the era before the GUT transition. 
The
system may actually
not be in thermodynamical equilibrium. This is the case in most of the
models of inflation.  The effects of the high curvature
of the universe should perhaps be taken into account
\cite{ref:Buccella}. Of course
the formalism can describe only systems in an equilibrium and therefore
it can be used only to study some of the features of the phase transition.
However, there are many important properties which can be extracted from
the equilibrium behavior near the transition. The effective 
three-dimensional theory gives a correct picture of the equilibrium
behaviour only at high temperatures and when the four-dimensional coupling
constants are small. These demands are well satisfied in SU(5). 

The previous numerical studies of the electroweak phase transition
have shown that for large values of the Higgs self-coupling constant
the transition ends. 
We expect the same phenomenon to occur also in the
case of SU(5) theory, but we have to wait for the numerical simulations
to confirm this conjecture. The absence of the transition may have some
important cosmological consequences.
It might affect the problem with the overabundance of magnetic monopoles.
It has been suggested \cite{ref:salomonson,ref:dvali,ref:bimonte} that
in the absence of a transition the monopole problem could be solved without
inflation.
The reliable non-perturbative
results of the SU(5) phase transition can also be used to make the
analysis of the possible inflationary phase of the evolution of the universe
much more accurate.

\section*{Acknowledgements}
I wish to thank K.~Kajantie, M.~Laine and M.~Shaposhnikov for useful
discussions. This work was supported by the Academy of Finland.

\appendix
\section{Properties of the SU(5) generators}
\label{app:SU5}
In the case of SU(5) symmetry, the structure of the vertices is much richer
than for example
in SU(2). Here is a somewhat complete list of the
relations needed in the
calculations of this paper. Some of the group-theoretical
factors necessary for the calculation of the 
effective potential must be evaluated explicitly by using a specific
choice of generators and we shall omit them here.
The relations of this appendix can be obtained using the methods
of Kaplan and Resnikoff \cite{ref:kaplan}. 

First we give some relations and conventions 
for the generators of the fundamental
representation of SU($N$), which are defined to be Hermitian:
\bea
[T^A,T^B]&=&if^{ABC}T^C,\\
\Tr T^AT^B&=&\frac{1}{2}\delta^{AB},\\
T^AT^B&=&\frac{1}{2N}\delta^{AB}+\frac{1}{2}d^{ABC}T^C+\frac{i}{2}f^{ABC}T^C,\\
(T^AT^A)_{ab}&=&\frac{N^2-1}{2N}\delta_{ab},\\
\Tr T^AT^BT^C&=&\frac{1}{4}(d^{ABC}+if^{ABC}),\\
\Tr T^AT^BT^CT^D&=&
\frac{1}{4N}\delta^{AB}\delta^{CD}+\frac{1}{8}(d^{ABE}+if^{ABE})
(d^{CDE}+if^{CDE}).
\eea

In order to discuss the 
properties of the generators of the adjoint representation, 
let us define two matrices
\beq
(F^A)_{BC}=-if^{ABC},\quad
(D^A)_{BC}=d^{ABC}.
\eeq
For these we have the usual Jacobi identities
\bea
f^{ABE}f^{CDE}+f^{CBE}f^{DAE}+f^{DBE}f^{ACE}&=&0,\\
f^{ABE}d^{CDE}+f^{CBE}d^{DAE}+f^{DBE}d^{ACE}&=&0.
\eea

Traces of two matrices:
\bea
\Tr F^AF^B&=&f^{ACD}f^{BCD}=N\delta^{AB},\\
\Tr D^AD^B&=&d^{ACD}d^{BCD}=\frac{N^2-4}{N}\delta^{AB}.
\eea

Traces of three matrices:
\bea
\Tr F^AF^BF^C&=&\frac{N}{2}if^{ABC},\\
\Tr F^AD^BD^C&=&\frac{N^2-4}{2N}if^{ABC},\\
\Tr D^AF^BF^C&=&\frac{N}{2}d^{ABC},\\
\Tr D^AD^BD^C&=&\frac{N^2-12}{2N}d^{ABC}.
\eea

Traces of four matrices:
\bea
\Tr F^AF^BF^CF^D
&=&\delta^{AB}\delta^{CD}+\delta^{AD}\delta^{BC}\nn
&&
+\frac{N}{4}(d^{ABE}d^{CDE}-d^{ACE}d^{BDE}+d^{ADE}d^{BCE}),\\
\Tr F^AF^BD^CD^D
&=&\delta^{AB}\delta^{CD}-\delta^{AD}\delta^{BC}\nn
&&
+\frac{N}{4}(d^{ABE}d^{CDE}+d^{ACE}d^{BDE}-d^{ADE}d^{BCE})\nn
&&
-\frac{N}{2}f^{ABE}f^{CDE}-\frac{2}{N}f^{ADE}f^{BCE},\\
\Tr D^AD^BD^CD^D
&=&\frac{N^2-8}{N^2}\delta^{AB}\delta^{CD}+\delta^{AD}\delta^{BC}
+\frac{N^2-24}{4N}d^{ABE}d^{CDE}\nn
&&-\frac{N}{4}d^{ACE}d^{BDE}
+\frac{N^2-8}{4N}d^{ADE}d^{BCE}\nn
&&
+\frac{2}{N}(f^{ABE}f^{CDE}+f^{ADE}f^{BCE}).
\eea

\section{Two-loop effective potential of the 3d theory}
\label{app:2loopep}
In this appendix we give the expressions for the effective potential of
the Higgs field in the three-dimensional theory broken to the two physically
most interesting directions $\tau_1$ and $\tau_2$.

When the symmetry is broken to the symmetry group 
SU(3)$\times$SU(2)$\times$U(1), $\Phi\mapsto\Phi+v\tau_1$, the effective potential is
\bea
V(v)&=&
\frac{1}{2}yv^2+\left(\frac{1}{4}x_1+\frac{7}{120}x_2\right)v^4\nn
&&+8C_\mathrm{S}(m_1)+12C_\mathrm{S}(m_2)+3C_\mathrm{S}(m_3)
+C_\mathrm{S}(m_4)+12C_\mathrm{V}(M)\nn
&&-\left(20x_1+10x_2\right)\dd{SS}(m_1,m_1)
-\left(42x_1+15x_2\right)\dd{SS}(m_2,m_2)\nn
&&
-\left(\frac{15}{4}x_1+\frac{15}{8}x_2\right)\dd{SS}(m_3,m_3)
-\left(\frac{3}{4}x_1+\frac{7}{40}x_2\right)\dd{SS}(m_4,m_4)\nn
&&
-\left(48x_1+16x_2\right)\dd{SS}(m_1,m_2)
-12x_2\dd{SS}(m_1,m_3)\nn
&&
-\left(4x_1+\frac{8}{5}x_2\right)\dd{SS}(m_1,m_4)
-\left(18x_1+9x_2\right)\dd{SS}(m_2,m_3)\nn
&&
-\left(6x_1+\frac{7}{5}x_2\right)\dd{SS}(m_2,m_4)
-\left(\frac{3}{2}x_1+\frac{27}{20}x_2\right)\dd{SS}(m_3,m_4)\nn
&&-6\dd{SV}(m_1,0)-\frac{15}{2}\dd{SV}(m_2,0)
-\frac{3}{2}\dd{SV}(m_3,0)-4\dd{SV}(m_1,M)\nn
&&-\frac{15}{2}\dd{SV}(m_2,M)-
\frac{9}{4}\dd{SV}(m_3,M)
-\frac{5}{4}\dd{SV}(m_4,M)\nn
&&-\frac{15}{4}\dd{VV}(0,0)-\frac{15}{2}\dd{VV}(M,0)-\frac{15}{4}
\dd{VV}(M,M)\nn
&&
-6\dd{SSV}(m_1,m_1,0)-\frac{15}{2}\dd{SSV}(m_2,m_2,0)
-\frac{3}{2}\dd{SSV}(m_3,m_3,0)\nn
&&-8\dd{SSV}(m_1,m_2,M)
-\frac{9}{2}\dd{SSV}(m_2,m_3,M)-\frac{5}{2}\dd{SSV}(m_2,m_4,M)\nn
&&-\frac{5}{2}\dd{VVV}(0,0,0)-\frac{15}{2}\dd{VVV}(0,M,M)\nn
&&-30\dd{\eta\eta V}(0)-30\dd{\eta\eta V}(M)\nn
&&-v^2\left(
 \frac{8}{3}x_2^2\dd{SSS}(m_1,m_1,m_1)
+8\left(x_1+\frac{2}{5}x_2\right)^2\dd{SSS}(m_1,m_1,m_4)\right.\nn
&&\phantom{-v^2}\left.
+\frac{4}{15}x_2^2\dd{SSS}(m_1,m_2,m_2)
+\frac{12}{5}x_2^2\dd{SSS}(m_2,m_2,m_3)\right.\nn
&&\phantom{-v^2}\left.
+12\left(x_1+\frac{7}{30}x_2\right)^2\dd{SSS}(m_2,m_2,m_4)\right.\nn
&&\phantom{-v^2}\left.
+3\left(x_1+\frac{9}{10}x_2\right)^2\dd{SSS}(m_3,m_3,m_4)\right.\nn
&&\phantom{-v^2}\left.
+3\left(x_1+\frac{7}{30}x_2\right)^2\dd{SSS}(m_4,m_4,m_4)\right.\nn
&&\phantom{-v^2}\left.
+\frac{5}{3}\dd{SVV}(m_1,M,M)+\frac{15}{16}\dd{SVV}(m_3,M,M)\right.\nn
&&\phantom{-v^2}\left.
+\frac{25}{28}\dd{SVV}(m_4,M,M)+\frac{25}{16}\dd{SVV}(m_2,0,M)
\right),
\eea
where the masses are
\bea
M^2&=&\frac{5}{12}v^2,\nn
m_1^2&=&y+\left(x_1+\frac{2}{5}x_2\right)v^2,\quad
m_2^2=y+\left(x_1+\frac{7}{30}x_2\right)v^2,\nn
m_3^2&=&y+\left(x_1+\frac{9}{10}x_2\right)v^2,\quad
m_4^2=y+\left(3x_1+\frac{7}{10}x_2\right)v^2.
\eea

When the shift is made to the direction $\Phi\mapsto\Phi+v\tau_2$,
the symmetry is broken down to SU(4)$\times$U(1) and the effective
potential becomes
\bea
V(v)&=&
\frac{1}{2}yv^2+\left(\frac{1}{4}x_1+\frac{13}{80}x_2\right)v^4\nn
&&+15C_\mathrm{S}(m_1)+8C_\mathrm{S}(m_2)+C_\mathrm{S}(m_3)
+8C_\mathrm{V}(M)\nn
&&-\left(\frac{255}{4}x_1+\frac{435}{16}x_2\right)\dd{SS}(m_1,m_1)
-\left(20x_1+10x_2\right)\dd{SS}(m_2,m_2)\nn
&&
-\left(\frac{3}{4}x_1+\frac{39}{80}x_2\right)\dd{SS}(m_3,m_3)
-\left(60x_1+15x_2\right)\dd{SS}(m_1,m_2)\nn
&&
-\left(\frac{15}{2}x_1+\frac{9}{8}x_2\right)\dd{SS}(m_1,m_3)
-\left(4x_1+\frac{13}{5}x_2\right)\dd{SS}(m_2,m_3)\nn
&&-15\dd{SV}(m_1,0)-5\dd{SV}(m_2,0)
-\frac{15}{4}\dd{SV}(m_1,M)\nn
&&-5\dd{SV}(m_2,M)-\frac{5}{4}\dd{SV}(m_3,M)\nn
&&-\frac{15}{2}\dd{VV}(0,0)-5\dd{VV}(M,0)-\frac{5}{2}
\dd{VV}(M,M)\nn
&&
-15\dd{SSV}(m_1,m_1,0)-5\dd{SSV}(m_2,m_2,0)\nn
&&
-\frac{15}{2}\dd{SSV}(m_1,m_2,M)
-\frac{5}{2}\dd{SSV}(m_2,m_3,M)\nn
&&-5\dd{VVV}(0,0,0)-5\dd{VVV}(0,M,M)\nn
&&-40\dd{\eta\eta V}(0)-20\dd{\eta\eta V}(M)\nn
&&-v^2\left(
 \frac{27}{8}x_2^2\dd{SSS}(m_1,m_1,m_1)\right.\nn
&&\phantom{-v^2}\left.
+15\left(x_1+\frac{3}{20}x_2\right)^2\dd{SSS}(m_1,m_1,m_3)
\right.\nn&&\phantom{-v^2}\left.
+\frac{3}{2}x_2^2\dd{SSS}(m_1,m_2,m_2)
+8\left(x_1+\frac{13}{20}x_2\right)^2\dd{SSS}(m_2,m_2,m_3)
\right.\nn&&\phantom{-v^2}\left.
+3\left(x_1+\frac{13}{20}x_2\right)^2\dd{SSS}(m_3,m_3,m_3)
\right.\nn&&\phantom{-v^2}\left.
+\frac{75}{32}\dd{SVV}(m_1,M,M)+\frac{25}{16}\dd{SVV}(m_2,0,M)
\right.\nn&&\phantom{-v^2}\left.
+\frac{25}{32}\dd{SVV}(m_3,M,M)
\right),
\eea
where the masses are
\bea
M^2&=&\frac{5}{8}v^2,\quad
m_1^2=y+\left(x_1+\frac{3}{20}x_2\right)v^2,\nn
m_2^2&=&y+\left(x_1+\frac{13}{20}x_2\right)v^2,\quad
m_3^2=y+\left(3x_1+\frac{39}{20}x_2\right)v^2.
\eea

The explicit expressions for the $C$ and $D$ functions are
\bea
C_\mathrm{S}(m)&=&-\frac{m^3}{12\pi},\nn
C_\mathrm{V}(m)&=&-\frac{m^3}{6\pi},\nn
\dd{SSS}(m_1,m_2,m_3)&=&
\frac{1}{16\pi^2}
\left(\frac{1}{2}+
\log\frac{\mu}{m_1+m_2+m_3}\right),\nn
\dd{SS}(m_1,m_2)&=&
-\frac{m_1m_2}{16\pi^2},\nn
\dd{SV}(m_1,m_2)&=&4\dd{SS}(m_1,m_2),\nn
\dd{VV}(m_1,m_2)&=&\frac{16}{3}\dd{SS}(m_1,m_2),\nn
\dd{SSV}(m_1,m_2,M)&=&
(M^2-2m_1^2-2m_2^2)\dd{SSS}(m_1,m_2,M)\nn
&&
+\frac{(m_1^2-m_2^2)^2}{M^2}
(\dd{SSS}(m_1,m_2,M)-\dd{SSS}(m_1,m_2,0))\nn
&&+
\frac{1}{16\pi^2M}((m_1+m_2)(M^2+(m_1-m_2)^2)\nn
&&-Mm_1m_2),\nn
\dd{\eta\eta V}(M)&=&-\frac{M^2}{4}\dd{SSS}(M,0,0)\nn
\dd{VVV}(0,M,M)&=&
\frac{1}{16\pi^2M^2}
\left(\frac{29}{12}
+8\log 2-10\log\frac{\mu}{M}\right)\nn
\dd{SVV}(m,0,M)&=&
\frac{1}{16\pi^2}
\left(\frac{3}{2}-\frac{2m}{M}
+\frac{2m^2}{M^2}\log\frac{m+M}{m}
+6\log\frac{u}{m+M}\right),\nn
\dd{SVV}(m,M,M)&=&
\frac{1}{16\pi^2}
\left(
3-\frac{2m}{M}-\frac{m^2}{M^2}
-\frac{m^4}{M^4}\log\frac{m}{m+2M}\right.\nn
&&\left.
+2\frac{(m^2-M^2)^2}{M^4}\log\frac{m+M}{m+2M}+6\log\frac{\mu}{m+2M}
\right).\nn
\eea

\end{fmffile}
\bibliographystyle{plain}
\bibliography{pap}
\end{document}

%% file: kuva1.tex
\begin{figure}
\begin{eqnarray}
\twopointblob{dashes}{(1.1)}(k)
&=&
\twopoint{dashes}{dashes}{photon}{(1.1.1)}.\nonumber\\
\twopointblob{dbl_wiggly}{(1.2)}(k)
&=&
\twopoint{dbl_wiggly}{photon}{photon}{(1.2.1)}+
\twopoint{dbl_wiggly}{dashes}{dashes}{(1.2.2)}+
\twopoint{dbl_wiggly}{double_arrow}{double_arrow}{(1.2.3)}.\nonumber\\
\twopointblob{photon}{(1.3)}(k)
&=&
\twopoint{photon}{photon}{photon}{(1.3.1)}+
\twopoint{photon}{dashes}{dashes}{(1.3.2)}+
\twopoint{photon}{double_arrow}{double_arrow}{(1.3.3)}.\nonumber\\
\twopointblob{dbl_wiggly}{(1.4)}(0)
&=&
\twopoint{dbl_wiggly}{photon}{photon}{(1.4.1)}+
\twopoint{dbl_wiggly}{dashes}{dashes}{(1.4.2)}+
\twopoint{dbl_wiggly}{double_arrow}{double_arrow}{(1.4.3)}\nonumber\\
&+&
\twopointtp{dbl_wiggly}{photon}{(1.4.4)}+
\twopointtp{dbl_wiggly}{dashes}{(1.4.5)}.\nonumber
\end{eqnarray}
\caption{The two-point diagrams needed in the super-heavy integration.
The dashed line corresponds to the Higgs field, the wavy line
to the gauge field spatial components, the double wavy line
to the gauge field temporal component and the double line to the 
ghost field.}
\label{fig:normitus}
\end{figure}

%% file: kuva2.tex
\begin{figure}
\begin{eqnarray}
\fourpointblob{dashes}{dashes}{photon}{photon}{(2.1)}&=&
\fourpointtrg{dashes}{dashes}{photon}{photon}{photon}{photon}{photon}{(2.1.1)}
+
\fourpointtrg{dashes}{dashes}{dashes}{dashes}{dashes}{photon}{photon}{(2.1.2)}
+
\fourpointfish{dashes}{dashes}{dashes}{dashes}{photon}{photon}{(2.1.3)}
\nonumber\\
&&+
\fourpointfish{dashes}{photon}{dashes}{photon}{dashes}{photon}{(2.1.4)}
+
\fourpointfish{dashes}{dashes}{photon}{photon}{photon}{photon}{(2.1.5)}
.\nonumber\\
\fourpointblob{dashes}{dashes}{dbl_wiggly}{dbl_wiggly}{(2.2)}&=&
\fourpointtrg{dashes}{dashes}{photon}{photon}{photon}{dbl_wiggly}{dbl_wiggly}
{(2.2.1)}+
\fourpointtrg{dashes}{dashes}{dashes}{dashes}{dashes}{dbl_wiggly}{dbl_wiggly}
{(2.2.2)}+
\fourpointfish{dashes}{dashes}{dashes}{dashes}{dbl_wiggly}{dbl_wiggly}
{(2.2.3)}\nonumber\\
&&+
\fourpointfish{dashes}{dbl_wiggly}{dashes}{photon}{dashes}{dbl_wiggly}
{(2.2.4)}+
\fourpointfish{dashes}{dashes}{photon}{photon}{dbl_wiggly}{dbl_wiggly}
{(2.2.5)}.
\nonumber\\
\fourpointblob{dbl_wiggly}{dbl_wiggly}{dbl_wiggly}{dbl_wiggly}{(2.3)}
&=&
\fourpointfish{dbl_wiggly}{dbl_wiggly}{dashes}{dashes}{dbl_wiggly}{dbl_wiggly}
{(2.3.1)}+
\fourpointfish{dbl_wiggly}{dbl_wiggly}{photon}{photon}{dbl_wiggly}{dbl_wiggly}
{(2.3.2)}+
\fourpointtrg{dbl_wiggly}{dbl_wiggly}{dashes}{dashes}{dashes}{dbl_wiggly}
{dbl_wiggly}{(2.3.3)}\nonumber\\&&+
\fourpointtrg{dbl_wiggly}{dbl_wiggly}{photon}{photon}{photon}{dbl_wiggly}
{dbl_wiggly}{(2.3.4)}+
\fourpointsqr{dbl_wiggly}{dbl_wiggly}{dashes}{dashes}{dashes}{dashes}
{dbl_wiggly}{dbl_wiggly}{(2.3.5)}+
\fourpointsqr{dbl_wiggly}{dbl_wiggly}{photon}{photon}{photon}{photon}
{dbl_wiggly}{dbl_wiggly}{(2.3.6)}\nonumber\\&&+
\fourpointsqr{dbl_wiggly}{dbl_wiggly}{double_arrow}{double_arrow}
{double_arrow}{double_arrow}{dbl_wiggly}{dbl_wiggly}{(2.3.7)}.\nonumber\\
\fourpointblob{dashes}{dashes}{dashes}{dashes}{(2.4)}
&=&
\fourpointfish{dashes}{dashes}{dashes}{dashes}{dashes}{dashes}{(2.4.1)}+
\fourpointfish{dashes}{dashes}{photon}{photon}{dashes}{dashes}{(2.4.2)}
.\nonumber\end{eqnarray}
\caption{The four-point diagrams needed in the super-heavy integration}
\label{fig:4point}
\end{figure}

%% file: kuva3.tex
\begin{figure}
$$
\begin{array}{cccc}
\vacuumo{dashes}{(3.1.1)}&
\vacuumo{photon}{(3.1.2)}&
\vacuumct{dashes}{(3.2.1)}&
\vacuumct{photon}{(3.2.2)}\\&&&\\
\vacuumeight{dashes}{dashes}{(3.3.1)}&
\vacuumeight{dashes}{photon}{(3.3.2)}&
\vacuumeight{photon}{photon}{(3.3.3)}&
\vacuumss{dashes}{photon}{dashes}{(3.3.4)}\\&&&\\
\vacuumss{photon}{photon}{photon}{(3.3.5)}&
\vacuumss{double_arrow}{photon}{double_arrow}{(3.3.6)}&
\vacuumss{dashes}{dashes}{dashes}{(3.3.7)}&
\vacuumss{photon}{dashes}{photon}{(3.3.8)}
\end{array}
$$
\caption{The diagrams for the effective potential of the Higgs field}
\label{fig:Efpot}
\end{figure}

%% file: kuva4.tex
\begin{figure}
$$
\begin{array}{cccc}
\twopoint{photon}{plain}{plain}{(4.1)}&
\fourpointfish{dashes}{dashes}{plain}{plain}{photon}{photon}{(4.2.1)}&
\fourpointtrg{dashes}{dashes}{plain}{plain}{plain}{photon}{photon}{(4.2.2)}&
\fourpointfish{dashes}{dashes}{plain}{plain}{dashes}{dashes}{(4.3)}\\&&&\\
\vacuumo{plain}{(4.4.1)}&
\vacuumeight{plain}{plain}{(4.4.2)}&
\vacuumeight{dashes}{plain}{(4.4.3)}&
\vacuumeight{photon}{plain}{(4.4.4)}\\&&&\\
\vacuumss{plain}{photon}{plain}{(4.4.5)}&
\vacuumss{plain}{dashes}{plain}{(4.4.6)}&&
\end{array}
$$
\caption{The diagrams needed for the heavy integration.
The solid line represents the $A_0$ scalar field.}
\label{fig:heavy}
\end{figure}